\documentclass[showpacs,showkeys,amsmath,amssymb,nofootinbib]{revtex4}%
\usepackage{graphicx}
\usepackage{epsfig}
\usepackage{bm}
\usepackage{amsmath}
\usepackage{amsfonts}
\usepackage{amssymb}
\usepackage{subfigure}%
\providecommand{\U}[1]{\protect\rule{.1in}{.1in}}

\makeatletter
\newcommand\erfc{\mathop{\operator@font erfc}\nolimits}
\def\slashchar#1{\setbox0=\hbox{$#1$}
\dimen0=\wd0 \setbox1=\hbox{/} \dimen1=\wd1
\ifdim\dimen0>\dimen1 \rlap{\hbox to \dimen0{\hfil/\hfil}} #1
\else  \rlap{\hbox to \dimen1{\hfil$#1$\hfil}} / \fi}

\makeatother
\begin{document}
\title{Rare decay $\pi^{0}\rightarrow e^{+}e^{-}$: theory confronts KTeV data.}
\author{Alexander E. Dorokhov, Mikhail A. Ivanov}
\affiliation{Joint Institute for Nuclear Research, Bogoliubov Laboratory of Theoretical
Physics, 114980, Moscow region, Dubna, Russia}
\begin{abstract}
Within the dispersive approach to the amplitude of the rare decay $\pi
^{0}\rightarrow e^{+}e^{-}$ the nontrivial dynamics is contained only in the
subtraction constant. We express this constant, in the leading order in
$\left(  m_{e}/\Lambda\right)  ^{2}$ perturbative series, in terms of the
inverse moment of the pion transition form factor given in symmetric
kinematics. By using the CELLO and CLEO data on the pion transition form
factor given in asymmetric kinematics the lower bound of the decay branching
ratio is found. The restrictions following from QCD allow us to make a
quantitative prediction for the branching $B\left(  \pi^{0}\rightarrow
e^{+}e^{-}\right)  =\left(  6.2\pm0.1\right)  \cdot10^{-8}$ which is $3\sigma$
below the recent KTeV measurement. We confirm our prediction by using the
quark models and phenomenological approaches based on the vector meson
dominance. The decays $\eta\rightarrow l^{+}l^{-}$ are also discussed.
\end{abstract}
\maketitle
1. Experimental measurements of neutral pseudoscalar meson decays into lepton
pairs and the comparison with theoretical predictions offer an interesting way
to study long-distance dynamics in the Standard Model. Recently, the KTeV
E799-II experiment at Fermilab has observed $\pi^{0}\rightarrow e^{+}e^{-}$
events using $K_{L}\rightarrow3\pi$ decay as a source of tagged neutral pions
\cite{Abouzaid:2007md}. The branching ratio of the pion decay into an
electron-positron pair was determined to be equal to
\begin{equation}
B^{\mathrm{KTeV}}\left(  \pi^{0}\rightarrow e^{+}e^{-},x_{D}>0.95\right)
=\left(  6.44\pm0.25\pm0.22\right)  \cdot10^{-8}, \label{KTeV0}%
\end{equation}
where $x_{D}\equiv\left(  m_{e^{+}e^{-}}/m_{\pi_{0}}\right)  ^{2}$ is the
Dalitz variable. By extrapolating the full radiative tail beyond $x_{D}>0.95$
and scaling the result back up by the overall radiative correction
\cite{Bergstrom:1982wk,Triantaphyllou:1993ww} to find the lowest-order rate
for $\pi^{0}\rightarrow e^{+}e^{-}$ \ the KTeV Collaboration obtained
\begin{equation}
B_{\mathrm{no-rad}}^{\mathrm{KTeV}}\left(  \pi^{0}\rightarrow e^{+}%
e^{-}\right)  =\left(  7.49\pm0.29\pm0.25\right)  \cdot10^{-8}. \label{KTeV}%
\end{equation}

The rare decay $\pi^{0}\rightarrow e^{+}e^{-}$ has been studied theoretically
over the years, starting with the first prediction of the rate by Drell
\cite{Drell59}. Since no spinless current coupling of quarks to leptons
exists, the decay is described in the lowest order of QED as a one-loop
process via the two-photon intermediate state, as shown in Fig. 1. A factor of
$2\left(  m_{e}/m_{\pi}\right)  ^{2}$ corresponding to the approximate
helicity conservation of the interaction and two orders of $\alpha$ suppress
the decay with respect to the $\pi^{0}\rightarrow\gamma\gamma$ decay, leading
to an expected branching ratio of about $10^{-7}$. In the Standard Model
contributions from the weak interaction to this process are many orders of
magnitude smaller and can be neglected. The interaction of leptons and quarks
with leptoquarks is a possible mechanism for the pion decay from physics
beyond the Standard Model. The confrontation of the theory and experiment will
have some influence on the problem of strong sector contribution to the muon
anomalous magnetic moment $g-2$ \cite{Dorokhov:2005ff,Passera:2004bj}.
\begin{figure}[th]
\includegraphics[width=5cm]{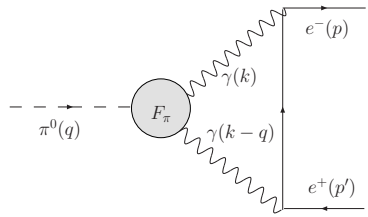}\caption{Triangle diagram for $\pi
^{0}\rightarrow e^{+}e^{-}$ process with a pion $\pi^{0}\rightarrow
\gamma^{\ast}\gamma^{\ast}$ form factor in the vertex.}%
\label{fig:triangle}%
\end{figure}

To lowest order in QED the normalized branching ratio is given by%
\begin{equation}
R\left(  \pi^{0}\rightarrow e^{+}e^{-}\right)  =\frac{B\left(  \pi
^{0}\rightarrow e^{+}e^{-}\right)  }{B\left(  \pi^{0}\rightarrow\gamma
\gamma\right)  }=2\left(  \frac{\alpha}{\pi}\frac{m_{e}}{m_{\pi}}\right)
^{2}\beta_{e}\left(  m_{\pi}^{2}\right)  \left\vert \mathcal{A}\left(  m_{\pi
}^{2}\right)  \right\vert ^{2}, \label{B}%
\end{equation}
where $\beta_{e}\left(  q^{2}\right)  =\sqrt{1-4\frac{m_{e}^{2}}{q^{2}}}$,
$B\left(  \pi^{0}\rightarrow\gamma\gamma\right)  =0.988$ \cite{Yao:2006px}. In
this article, we describe the process $\pi^{0}\rightarrow e^{+}e^{-}$ by the
diagram shown in Fig.1, where $F_{\pi\gamma^{\ast}\gamma^{\ast}}$ is the form
factor of the transition $\pi^{0}\rightarrow\gamma^{\ast}\gamma^{\ast}$ with
off-shell photons. The reduced amplitude $\mathcal{A}$ can be written as%
\begin{equation}
\mathcal{A}\left(  q^{2}\right)  =\frac{2i}{q^{2}}\int\frac{d^{4}k}{\pi^{2}%
}\frac{q^{2}k^{2}-\left(  qk\right)  ^{2}}{\left(  k^{2}+i\varepsilon\right)
\left(  \left(  k-q\right)  ^{2}+i\varepsilon\right)  \left(  \left(
k-p\right)  ^{2}-m_{e}^{2}+i\varepsilon\right)  }F_{\pi\gamma^{\ast}%
\gamma^{\ast}}\left(  -k^{2},-\left(  k-q\right)  ^{2}\right)  ,\label{R}%
\end{equation}
where $q^{2}=m_{\pi}^{2},p^{2}=m_{e}^{2}$. We put the sign minus in the
arguments of the form factor explicitly to emphasize that Eq.~(\ref{R}) is
written in the Minkowski space. The form factor is normalized as $F_{\pi
\gamma^{\ast}\gamma^{\ast}}(0,0)=1$ and falls down quite rapidly in the
Euclidean region of momenta to provide the ultraviolet convergence of the
integral. A number of model calculations of the amplitude $\mathcal{A}\left(
q^{2}\right)  $ was performed
\cite{Drell59,Berman60,Quigg68,Efimov:1981vh,Bergstrom:1982zq,
Bergstrom:1983ay,Silagadze:2006rt} by employing different shapes of the form
factor $F_{\pi\gamma^{\ast}\gamma^{\ast}}$. We discuss some of them below.

The aim of the present paper is to calculate the branching ratio
$B\left(  \pi^{0}\rightarrow e^{+}e^{-}\right)$
and estimate the uncertainties by using the available experimental and
theoretical information on the pion transition form factor. In particular, the
important constraints follow from the results obtained by the CELLO and CLEO
collaborations and restrictions set by QCD.

2. First, we derive a suitable representation for the amplitude in
Eq.~(\ref{R}) which would help us to perform a straightforward analysis by
using the available information on the pion transition form factor. To do
this, we employ the dispersive approach to the calculation of the amplitude
developed in many papers (see, e.g. \cite{Bergstrom:1983ay} and references
therein). The imaginary part of the amplitude in Eq.~(\ref{R})
\begin{equation}
\operatorname{Im}\mathcal{A}\left(  q^{2}\right)  =\frac{\pi}{2\beta
_{e}\left(  q^{2}\right)  }\ln\left(  y_{e}\left(  q^{2}\right)  \right)
,\qquad y_{e}\left(  q^{2}\right)  =\frac{1-\beta_{e}\left(  q^{2}\right)
}{1+\beta_{e}\left(  q^{2}\right)  },\label{Im}%
\end{equation}
comes from the contribution of real photons in the intermediate state and is
model independent since $F_{\pi\gamma^{\ast}\gamma^{\ast}}\left(  0,0\right)
=1$. Using $\left\vert \mathcal{A}\right\vert ^{2}\geq\left(
\operatorname{Im}\mathcal{A}\right)  ^{2}$ and neglecting radiative
corrections one can get the well-known unitary bound for the branching ratio
in Eq.~(\ref{B}) \cite{Berman60}%
\begin{equation}
B\left(  \pi^{0}\rightarrow e^{+}e^{-}\right)  \geq B^{\mathrm{unitary}%
}\left(  \pi^{0}\rightarrow e^{+}e^{-}\right)  =4.69\cdot10^{-8}.\label{UnitB}%
\end{equation}

A once-subtracted dispersion relation for the amplitude in Eq.~(\ref{R}) is
written as \cite{Bergstrom:1983ay}
\begin{equation}
\mathcal{A}\left(  q^{2}\right)  =\mathcal{A}\left(  q^{2}=0\right)
+\frac{q^{2}}{\pi}\int_{0}^{\infty}ds\frac{\operatorname{Im}\mathcal{A}\left(
s\right)  }{s\left(  s-q^{2}\right)  }.\label{DispRel}%
\end{equation}
The second term in Eq.~(\ref{DispRel}) takes into account strong $q^{2}$
dependence of the amplitude around the point $q^{2}=0$ occurring due to the
branch cut coming from the two-photon intermediate state. Integrating
Eq.~(\ref{DispRel}) one arrives for $q^{2}\geq4m_{e}^{2}$ at
\cite{D'Ambrosio:1986ze,Savage:1992ac,Ametller:1993we}%
\begin{equation}
\operatorname{Re}\mathcal{A}\left(  q^{2}\right)  =\mathcal{A}\left(
q^{2}=0\right)  +\frac{1}{\beta_{e}\left(  q^{2}\right)  }\left[  \frac{1}%
{4}\ln^{2}\left(  y_{e}\left(  q^{2}\right)  \right)  +\frac{\pi^{2}}%
{12}+\mathrm{Li}_{2}\left(  -y_{e}\left(  q^{2}\right)  \right)  \right]
,\label{Rqb}%
\end{equation}
where $\mathrm{Li}_{2}\left(  z\right)  =-\int_{0}^{z}\left(  dt/t\right)
\ln\left(  1-t\right)  $ is the dilogarithm function\footnote{For completeness
we give explicit expressions for the amplitude $\tilde{\mathcal{A}}\left(
q^{2}\right)  =\mathcal{A}\left(  q^{2}\right)  -\mathcal{A}\left(  0\right)
$ for different regions of $q^{2}$: $\operatorname{Re}\tilde{\mathcal{A}%
}(q^{2})=\frac{1}{\beta(q^{2})}\Big[\mathrm{Li}_{2}(-y\left(  q^{2}\right)
)+\frac{\pi^{2}}{3}+\frac{1}{4}\ln^{2}(-y\left(  q^{2}\right)  )\Big],\hspace
{0.5cm}\operatorname{Im}\tilde{\mathcal{A}}(q^{2})=0,\,\,\,\mathrm{for}%
\,\,\,q^{2}\leq0;$ and $\operatorname{Re}\tilde{\mathcal{A}}(q^{2})=-\frac
{1}{\tilde{\beta}(q^{2})}\,\mathrm{Cl}_{2}(-2\theta),\hspace{0.5cm}%
\operatorname{Im}\tilde{\mathcal{A}}(q^{2})=-\frac{\pi}{\tilde{\beta}(q^{2}%
)}\,\mathrm{arctg}\Big[\tilde{\beta}(q^{2})\Big],\hspace{0.5cm}\mathrm{for}%
\,\,\,0\leq q^{2}\leq4m^{2}$. Here $\beta(q^{2})=\sqrt{1-4m^{2}/q^{2}}$,
$\tilde{\beta}(q^{2})=\sqrt{4m^{2}/q^{2}-1}$, $\theta=\mathrm{arctg}%
\Big[1/\tilde{\beta}(q^{2})\Big]$, and $\mathrm{Cl}_{2}(z)=-\int
\limits_{0}^{z}dt\ln|2\sin(t/2)|$ is the Clausen's integral.}. For the pion in
the leading order in $\left(  m_{e}/m_{\pi}\right)  ^{2},$ one gets
\begin{equation}
\operatorname{Re}\mathcal{A}\left(  m_{\pi}^{2}\right)  =\mathcal{A}\left(
q^{2}=0\right)  +\ln^{2}\left(  \frac{m_{e}}{m_{\pi}}\right)  +\frac{\pi^{2}%
}{12}.
\end{equation}

Thus, the nontrivial dynamics is only contained in the subtraction constant
$\mathcal{A}\left(  q^{2}=0\right)  $. We evaluate this quantity in the
following way \cite{Efimov:1981vh}. We use the double Mellin representation
for the pion transition form factor reducing the integral in Eq.~(\ref{R}) to
the convolution of propagator-like expressions. Then we perform the loop
integration by using the standard Feynman $\alpha$-representation. Finally, we
are able to expand the integral over the ratios of the electron and pion
masses to the characteristic scale of the pion form factor $\Lambda$ by
closing the Mellin contours in the appropriate manner and take the leading
term of expansion. We arrive at the following representation:
\begin{equation}
\mathcal{A}\left(  q^{2}=0\right)  =3\ln\left(  \frac{m_{e}}{\mu}\right)
+\chi_{P}\left(  \mu\right)  ,\label{RM0}%
\end{equation}
where the constant $\chi_{P}\left(  \mu\right)  $ is defined by
\begin{equation}
\chi_{P}\left(  \mu\right)  =-\frac{5}{4}+\frac{3}{2}\int_{0}^{\infty}%
dt\ln\left(  \frac{t}{\mu^{2}}\right)  \frac{\partial F_{\pi\gamma^{\ast
}\gamma^{\ast}}\left(  t,t\right)  }{\partial t}=-\frac{5}{4}-\frac{3}%
{2}\left[  \int_{0}^{\mu^{2}}dt\frac{F_{\pi\gamma^{\ast}\gamma^{\ast}}\left(
t,t\right)  -1}{t}+\int_{\mu^{2}}^{\infty}dt\frac{F_{\pi\gamma^{\ast}%
\gamma^{\ast}}\left(  t,t\right)  }{t}\right]  ,\label{RM2}%
\end{equation}
with $F_{\pi\gamma^{\ast}\gamma^{\ast}}\left(  t,t\right)  $ being the
physical pion transition form factor given in symmetric kinematics for
space-like photon momenta $t=Q^{2}=-q^{2}>0.$ One has to note that the
logarithmic dependence on the scale $\mu$ appearing in Eq.~(\ref{RM0}) as a
result of the decomposition of the integral over the dimensional variable $t$
into two parts is compensated by the scale dependence of the low-energy
constant $\chi_{P}\left(  \mu\right)  $ displayed in Eq.~(\ref{RM2}). The
obtained representation defines the unknown subtraction constant in dispersion
formula, Eq.~(\ref{DispRel}), via the pion transition form factor in a simple
and transparent way. It is consistent with the result of the chiral
perturbation theory
\cite{D'Ambrosio:1986ze,Savage:1992ac,Ametller:1993we,GomezDumm:1998gw,Knecht:1999gb}
where $\chi_{P}\left(  \mu\right)  $ is the unknown low-energy constant. The
result of Eq.~(\ref{RM0}) also agrees with the conclusions made in
\cite{Bergstrom:1983ay} where the inequalities $m_{e}<<m_{\pi}<<\Lambda$ were exploited.

3. In order to estimate the integral in Eq.~(\ref{RM2}), one needs to define
the pion transition form factor in symmetric kinematics for space-like photon
momenta. Since it is not known from the first principles, we will adapt the
available experimental data to perform such estimates. Let us first use the
fact that $F_{\pi\gamma^{\ast}\gamma^{\ast}}\left(  t,t\right)  <F_{\pi
\gamma^{\ast}\gamma^{\ast}}\left(  t,0\right)  $ for $t>0$ in order to obtain
the lower bound of the integral in Eq.~(\ref{RM2}). For this purpose, we take
the experimental results from the CELLO \cite{Behrend:1990sr} and CLEO
\cite{Gronberg:1997fj} Collaborations for the pion transition form factor in
asymmetric kinematics for space-like photon momentum which is well
parametrized by the monopole form
\begin{equation}
F_{\pi\gamma^{\ast}\gamma^{\ast}}^{\mathrm{CLEO}}\left(  t,0\right)  =\frac
{1}{1+t/s_{0}^{\mathrm{CLEO}}},\qquad s_{0}^{\mathrm{CLEO}}=\left(
776\pm22\quad\mathrm{MeV}\right)  ^{2}.\label{VMD1}%
\end{equation}
Note that $\mathrm{s_{0}^{CLEO}\approx}m_{\rho}^{2},$ as predicted by simple
vector meson dominance models (VMD) \cite{Berman60,Quigg68,Bergstrom:1983ay},
and is not far from the asymptotic prediction of operator product expansion
(OPE) QCD \cite{Lepage:1980fj}: $s_{0}^{OPE}=8\pi^{2}f_{\pi}^{2}=(821$
MeV)$^{2}$. For this type of the form factor one finds from Eqs.~(\ref{RM0})
and (\ref{RM2}) that
\begin{equation}
\mathcal{A}\left(  q^{2}=0\right)  >-\frac{3}{2}\ln\left(  \frac
{s_{0}^{\mathrm{CLEO}}}{m_{e}^{2}}\right)  -\frac{5}{4}=-23.2\pm
0.1.\label{Rcleo}%
\end{equation}
Thus, for the branching ratio we are able to establish the important lower
bound which considerably improves the unitary bound given by Eq.~(\ref{UnitB}%
)
\begin{equation}
B\left(  \pi^{0}\rightarrow e^{+}e^{-}\right)  >B^{\mathrm{CLEO}}\left(
\pi^{0}\rightarrow e^{+}e^{-}\right)  =\left(  5.84\pm0.02\right)
\cdot10^{-8}.
\end{equation}

Now let us proceed further in this manner and assume that the monopole form is
also a good parametrization for the form factor in symmetric kinematics
\begin{equation}
F_{\pi\gamma^{\ast}\gamma^{\ast}}\left(  t,t\right)  =\frac{1}{1+t/s_{1}%
}.\label{Ftt}%
\end{equation}
The scale $s_{1}$ can be fixed from the relation for the slopes of the form
factors in symmetric and asymmetric kinematics at low $t$
\cite{Dorokhov:2003sc},
\begin{equation}
\left.  -\frac{\partial F_{\pi\gamma^{\ast}\gamma^{\ast}}\left(  t,t\right)
}{\partial t}\right\vert _{t=0}=\left.  -2\frac{\partial F_{\pi\gamma^{\ast
}\gamma^{\ast}}\left(  t,0\right)  }{\partial t}\right\vert _{t=0}%
,\label{slope}%
\end{equation}
that gives $s_{1}=s_{0}/2$. Note that similar reduction of the scale is also
predicted by OPE QCD from the large momentum behavior of the form factors:
$s_{1}^{OPE}=s_{0}^{OPE}/3$ \cite{Lepage:1980fj}. Thus, the estimate for the
amplitude in the limit $q^{2}\rightarrow0$ can be obtained from
Eq.~(\ref{Rcleo}) by shifting the lower bound by a positive number which
belongs to the interval $[3\ln(2)/2,3\ln(3)/2]$. One finds
\begin{equation}
\mathcal{A}\left(  q^{2}=0\right)  =-\frac{3}{2}\ln\left(  \frac{s_{1}}%
{m_{e}^{2}}\right)  -\frac{5}{4}=-21.9\pm0.3,\label{R0t}%
\end{equation}
that corresponds to the value $\chi_{P}\left(  m_{\rho}\right)  =0.1\pm0.3$ of
the low-energy constant $\chi_{P}$ taken at the scale $\mu=m_{\rho}$. Using
the obtained prediction for the subtraction constant, one can evaluate the
branching ratio
\begin{equation}
B\left(  \pi^{0}\rightarrow e^{+}e^{-}\right)  =\left(  6.23\pm0.09\right)
\cdot10^{-8}.\label{Bt}%
\end{equation}
This is $3$ standard deviations lower than the KTeV result given by
Eq.~(\ref{KTeV}). One can convert the experimental data to obtain the
restriction on the scale parameter $s_{1}$ in Eq.~(\ref{R0t}). Then for the
amplitude at $q^{2}=0$ estimated from the experimental data for
$B_{\mathrm{no-rad}}^{\mathrm{KTeV}}\left(  \pi^{0}\rightarrow e^{+}%
e^{-}\right)  $ one finds
\begin{equation}
\mathcal{A}^{\mathrm{KTeV}}\left(  q^{2}=0\right)  =-18.6\pm0.9.\label{R0}%
\end{equation}
Since the amplitude in Eq.~(\ref{R0t}) depends logarithmically on the scale
parameter $s_{1}$, one needs to reduce the value of $s_{1}$ by a factor of
larger than $4$. However, it obviously contradicts the experimental data for
the slope parameters discussed above.

4. Let us compare our estimates with the results obtained in other approaches.
We start with the QCD sum rule approach \cite{Nesterenko:1982dn}. There, the
pion form factor of the transition process $\gamma^{\ast}\gamma^{\ast
}\rightarrow\pi^{0}$ was found in the form
\begin{equation}
F_{\pi\gamma^{\ast}\gamma^{\ast}}^{\mathrm{QCDsr}}\left(  t,t\right)
=2\int_{0}^{s_{0}^{\mathrm{QCDsr}}}ds\int_{0}^{1}dx\frac{x\left(  1-x\right)
t^{2}}{\left[  x\left(  1-x\right)  s+t\right]  ^{3}}+\mathrm{v.c.,}%
\label{Frad}%
\end{equation}
where $\mathrm{v.c.}$ are small corrections from the vacuum condensates and
$s_{0}^{\mathrm{QCDsr}}$ is the so-called dual interval parameter taken as
$s_{0}^{\mathrm{QCDsr}}=s_{0}^{\mathrm{OPE}}=0.7$ GeV$^{2}$ in the original
work \cite{Nesterenko:1982dn}. By using the relation for the form factor
slopes in Eq.~(\ref{slope}) and the expression for the radii given by
\begin{equation}
\left\langle r^{2}\right\rangle _{\pi^{0}\gamma^{\ast}\gamma^{\ast}%
}^{\mathrm{QCDsr}}=-6\left.  \frac{\partial F_{\pi\gamma^{\ast}\gamma^{\ast}%
}\left(  t,t\right)  }{\partial t}\right\vert _{t=0}=\frac{12}{s_{0}%
^{\mathrm{QCDsr}}},
\end{equation}
one can identify the duality parameter with the CLEO parameter
\begin{equation}
s_{0}^{\mathrm{QCDsr}}=s_{0}^{\mathrm{CLEO}}.
\end{equation}
Then by using Eq.~(\ref{RM2}) one finds
\begin{equation}
\mathcal{A}^{\mathrm{QCDsr}}\left(  q^{2}=0\right)  =-\frac{3}{2}\ln\left(
\frac{s_{0}^{\mathrm{CLEO}}}{m_{e}^{2}}\right)  +\frac{1}{4}=-21.7\pm
0.1,\label{Rsr}%
\end{equation}
that corresponds to the rescaling $s_{1}^{\mathrm{QCDsr}}=\left(
s_{0}^{\mathrm{QCDsr}}/e\right)  $ and is well suited to the interval in
Eq.~(\ref{R0t}). The corresponding branching ratio is shown in Table 1.
\begin{table}[th]
\caption[Results]{Values of the quantity $\mathcal{A}\left(  q^{2}=0\right)  $
and the branching ratio $B\left(  \pi^{0}\rightarrow e^{+}e^{-}\right)  $
obtained in our approach (see, Eq~(\ref{R0t})) and compared with various
phenomenological models and the KTeV experimental result.}%
\begin{tabular}
[c]{|c|c|c|c|c|c|c|c|}\hline
& CLEO+OPE & QCDsr & gVMD & QM\cite{Bergstrom:1983ay} & N$\chi$QM &
NQM \cite{Efimov:1981vh} & Experiment\cite{Abouzaid:2007md}\\\hline
$-\mathcal{A}\left(  q^{2}=0\right)  $ & $21.9\pm0.3$ & $21.7\pm0.1$ & $21.9$
& $23.4\pm0.5$ & $22.1\pm0.5$ & $24.5$ & $18.6\pm0.9$\\\hline
$B\left(  \pi^{0}\rightarrow e^{+}e^{-}\right)  \times10^{8}$ & $6.23\pm0.09$
& $6.21\pm0.05$ & $6.2$ & $5.8\pm0.2$ & $6.1\pm0.2$ & $5.38$ & $7.49\pm
0.38$\\\hline
\end{tabular}
\label{table1}%
\end{table}
%\bigskip

5. Then, we consider the parametrization of the pion form factor motivated by
the generalized VMD \cite{Knecht:2001xc}:
\begin{equation}
F_{\pi\gamma^{\ast}\gamma^{\ast}}^{\mathrm{gVMD}}\left(  s,t\right)
=\frac{4\pi^{2}f_{\pi}^{2}}{3}\frac{\left(  s+t\right)  st-h_{2}%
st+h_{5}\left(  s+t\right)  +M_{V}^{4}M_{V1}^{4}h_{7}}{\left(  M_{V}%
^{2}+s\right)  \left(  M_{V}^{2}+t\right)  \left(  M_{V1}^{2}+s\right)
\left(  M_{V1}^{2}+t\right)  }\label{gVMD}%
\end{equation}
with the parameters $M_{V}=769$ MeV, $M_{V1}=1465$ MeV, $h_{2}=-10$ GeV$^{2}$,
$h_{5}=6.93$ GeV$^{4}$, $h_{7}=3/\left(  4\pi^{2}f_{\pi}^{2}\right)  $. This
parametrization satisfies above mentioned constraints on the pion transition
form factor. At zero virtualities, the form factor is normalized by the axial
anomaly. The relation in Eq.~(\ref{slope}) is valid, which yields for the
radius
\begin{equation}
\left\langle r^{2}\right\rangle _{\pi\gamma\gamma^{\ast}}^{\mathrm{gVMD}%
}=-6\left.  \frac{\partial F\left(  t,0\right)  }{\partial t}\right\vert
_{t=0}=6\left(  \frac{1}{M_{V}^{2}}+\frac{1}{M_{V1}^{2}}-\frac{h_{5}}%
{M_{V}^{2}M_{V1}^{2}}\frac{3}{4\pi^{2}f_{\pi}^{2}}\right)  =0.39\quad
\mathrm{fm}^{2}\label{Rpigg}%
\end{equation}
which is close to PDG\ average $\left\langle r^{2}\right\rangle _{\pi
\gamma\gamma^{\ast}}^{\mathrm{PDG}}=0.407\pm0.051\quad\mathrm{fm}^{2}$
\cite{Yao:2006px}. Note also that numerically the second and third terms in
Eq.~(\ref{Rpigg}) almost cancel each other. The form factor $F_{\pi
\gamma^{\ast}\gamma^{\ast}}^{\mathrm{gVMD}}\left(  s,t\right)  $ has also
correct OPE QCD\ motivated behavior at large virtualities \cite{Lepage:1980fj}%
\begin{align}
\left.  F_{\pi\gamma^{\ast}\gamma^{\ast}}^{\mathrm{OPE}}\left(  t,0\right)
\right\vert _{t\rightarrow\infty} &  =\frac{4\pi^{2}f_{\pi}^{2}}{3}\frac
{h_{5}}{M_{V}^{2}M_{V1}^{2}}\frac{1}{t},\label{Fk0}\\
\left.  F_{\pi\gamma^{\ast}\gamma^{\ast}}^{\mathrm{OPE}}\left(  t,t\right)
\right\vert _{t\rightarrow\infty} &  =\frac{8\pi^{2}f_{\pi}^{2}}{3}\frac{1}%
{t}.\label{Fkq}%
\end{align}
The constant $h_{5}$ was fixed in \cite{Knecht:2001xc} from a fit of
CLEO\ data \cite{Gronberg:1997fj}, the constant $h_{2}$ defining the
next-to-leading order power correction to the form factor at large $t$ was
estimated in \cite{Melnikov:2003xd} using the results of QCD sum rules
\cite{Novikov:1983jt}. In the asymptotic OPE\ QCD limit, one has
$h_{5}/\left(  M_{V}^{2}M_{V1}^{2}\right)  \rightarrow6$.

For the generalized VMD form factor we estimate the subtraction constant
\begin{equation}
\mathcal{A}^{g\mathrm{VMD}}\left(  q^{2}=0\right)  =-3\ln\left(  \frac{M_{V}%
}{m_{e}}\right)  +\frac{1}{4}+\frac{3r}{\left(  r-1\right)  ^{2}}-\frac{3}%
{2}\frac{3r-1}{\left(  r-1\right)  ^{3}}\ln r+\nonumber
\end{equation}%
\begin{equation}
+\frac{4\pi^{2}f_{\pi}^{2}}{M_{V}^{2}\left(  r-1\right)  ^{3}}\left(
\frac{h_{2}}{2M_{V}^{2}}\left(  \left(  r+1\right)  \ln r-2\left(  r-1\right)
\right)  -\left(  1+\frac{h_{5}}{M_{V1}^{2}M_{V}^{2}}\right)  \left(
r^{2}-1-2r\ln r\right)  \right)  =-21.94,
\end{equation}
where $r=\left(  M_{V1}/M_{V}\right)  ^{2}$. It agrees well with our
prediction given by Eq.~(\ref{R0t}).

6. Let us now consider the amplitude $\pi^{0}\rightarrow e^{+}e^{-}$ in the
context of the constituent quark models. Within this model, the pion form
factor is given by the quark-loop (triangle) diagram. Taking the constituent
quark mass in the loop to be momentum independent, the result for the form
factor in symmetric kinematics is given by \cite{Ametller:1983ec}%
\begin{equation}
F_{\pi\gamma^{\ast}\gamma^{\ast}}^{\mathrm{QM}}\left(  t,t\right)
=\frac{2M_{q}^{2}}{\beta_{q}\left(  t\right)  t}\ln\left(  \frac{\beta
_{q}\left(  t\right)  +1}{\beta_{q}\left(  t\right)  -1}\right)  ,\qquad
\beta_{q}\left(  t\right)  =\sqrt{1+4\frac{M_{q}^{2}}{t}}.\label{Fqtt}%
\end{equation}
Substituting it into Eq.~(\ref{RM2}), one can get
\begin{equation}
\mathcal{A}_{\mathrm{QM}}\left(  q^{2}=0\right)  =3\ln\left(  \frac{m_{e}%
}{M_{q}}\right)  -\frac{17}{4},
\end{equation}
in accordance with \cite{Bergstrom:1983ay}. For the constituent quark mass in
the interval $M_{q}=300\pm50$ MeV one finds%
\begin{equation}
\mathcal{A}_{\mathrm{QM}}\left(  q^{2}=0\right)  =-\left(  23.4\pm0.5\right)
,
\end{equation}
which is in agreement with the above estimates but contradicts the
experimental result. In order to fit the KTeV value in Eq.~(\ref{R0}), one
needs to take the quark mass $M_{q}<100$ MeV, which is an unacceptable region
for the constituent quark mass.

By using the results for the pion transition form factor
\cite{Dorokhov:2002iu} obtained in the nonlocal quark model based on the
instanton picture of QCD vacuum \cite{Anikin:2000rq}, one finds numerically
from Eq.~(\ref{RM0}) that
\begin{equation}
\mathcal{A}_{\chi\mathrm{QM}}\left(  q^{2}=0\right)  =-\left(  22.1\pm
0.5\right)  ,
\end{equation}
which is consistent with Eq.~(\ref{R0t}). Within the instanton model, the
quark mass is momentum dependent and for simplicity is taken in the Gaussian
form
\begin{equation}
M\left(  k\right)  =M_{q}\exp\left(  -k^{2}/\Lambda_{\chi\mathrm{QM}}%
^{2}\right)  ,
\end{equation}
with $M_{q}=300\pm50$ MeV and $\Lambda_{\chi\mathrm{QM}}\sim1$ GeV fixed by
fitting the pion decay constant $f_{\pi}$.

The similar result was obtained in \cite{Efimov:1981vh} in the nonlocal quark model (NQM)
\cite{Efimov:1988yd,Efimov:1993zg} (see, Table 1). This model is based on the assumption
that the quark propagator is described by the entire function without any singularities
in the momentum space. Such an assumption guarantees the quark confinement in a sense
that eliminates the threshold cuts corresponding to free quark production.
A unified and uniformly accurate description of broad range of physical observables
is obtained within the framework of this model, see, e.g. \cite{Efimov:1993zg}.

One has to note that
the nonlocal quark model with momentum dependent mass has an important
advantage as compared to the mass independent (local) quark model. The first
model has correct large momentum behavior given by Eq.~(\ref{Fkq}), while the
latter has an extra $\log\left(  t\right)  $ term in the asymptotic, as
follows from Eq.~(\ref{Fqtt}). For this reason, the local model is not
suitable for fitting the CLEO data on the pion transition form factor.

7. The $\eta\rightarrow l^{+}l^{-}$ decay can be analyzed in a similar manner.
As in the pion case, the CLEO Collaboration has parametrized the data for the
$\eta$-meson in the monopole form \cite{Gronberg:1997fj}:
\begin{equation}
F_{\eta\gamma^{\ast}\gamma^{\ast}}^{\mathrm{CLEO}}\left(  t,0\right)
=\frac{1}{1+t/s_{0\eta}^{\mathrm{CLEO}}},\qquad s_{0\eta}^{\mathrm{CLEO}%
}=\left(  774\pm29\quad\mathrm{MeV}\right)  ^{2},
\end{equation}
which is very close to the relevant pion parameter. Then following the pion
case (with evident substitutions), one finds the bounds for the $q^{2}%
\rightarrow0$ limit of the amplitude $\eta\rightarrow\mu^{+}\mu^{-}$ as
\begin{equation}
\mathcal{A}_{\eta}\left(  q^{2}=0\right)  >-\frac{3}{2}\ln\left(
\frac{s_{0\eta}^{\mathrm{CLEO}}}{m_{\mu}^{2}}\right)  -\frac{5}{4}=-\left(
7.2\pm0.1\right)  ,
\end{equation}
and for $\eta\rightarrow e^{+}e^{-}$ one gets again Eq.~(\ref{Rcleo}). The
obtained estimates allow one to find the bounds for the branching ratios
\begin{align}
B\left(  \eta\rightarrow\mu^{+}\mu^{-}\right)   &  <\left(  6.23\pm
0.12\right)  \cdot10^{-6},\label{RbEta}\\
B\left(  \eta\rightarrow e^{+}e^{-}\right)   &  >\left(  4.33\pm0.02\right)
\cdot10^{-9}.\nonumber
\end{align}
Note that for the decay $\eta\rightarrow\mu^{+}\mu^{-}$ we get the upper limit
for the branching. This is because the real part of the amplitude for this
process taken at the physical point $q^{2}=m_{\eta}^{2}$ for the parameter
$s_{0\eta}^{\mathrm{CLEO}}$ remains negative and a positive shift due to the
change of the scale $s_{0\eta}\rightarrow s_{1\eta}$ reduces the absolute
value of the real part of the amplitude $\left\vert \operatorname{Re}%
\mathcal{A}\left(  m_{\eta}^{2}\right)  \right\vert $. At the same time,
considering the decays of $\pi^{0}$ and $\eta$ into an electron-positron pair,
the evolution to physical point (\ref{Rqb}) makes the real part of the
amplitude to be positive for the parameter $s_{0}^{\mathrm{CLEO}}$ and the
absolute value of the real part of the amplitude increases in changing the
scales of the meson form factors. The predictions for the decays
$\eta\rightarrow l^{+}l^{-}$ obtained by reducing the scale $s_{0\eta
}^{\mathrm{CLEO}}\rightarrow s_{1\eta}$ for the case of the $\eta$-meson
transition form factor are given in Table II. \begin{table}[th]
\caption[Results]{Values of the branchings $B\left(  P\rightarrow l^{+}%
l^{-}\right)  $ obtained in our approach and compared with the available
experimental results. }%
\begin{tabular}
[c]{|c|c|c|c|c|}\hline
$B$ & Unitary bound & CLEO bound & CLEO+OPE & Experiment\\\hline
$B\left(  \pi^{0}\rightarrow e^{+}e^{-}\right)  \times10^{8}$ & $\geq4.69$ &
$\geq5.85\pm0.03$ & $6.23\pm0.09$ & $7.49\pm0.38$ \cite{Abouzaid:2007md}%
\\\hline
$B\left(  \eta\rightarrow\mu^{+}\mu^{-}\right)  \times10^{6}$ & $\geq4.36$ &
$\leq6.23\pm0.12$ & $5.11\pm0.20$ & $5.8\pm0.8$ \cite{Yao:2006px,Abegg:1994wx}%
\\\hline
$B\left(  \eta\rightarrow e^{+}e^{-}\right)  \times10^{9}$ & $\geq1.78$ &
$\geq4.33\pm0.02$ & $4.60\pm0.06$ & $...$\\\hline
\end{tabular}
\label{table2}%
\end{table}

8. In this work, we have derived in the leading order in $\left(
m_{e}/\Lambda\right)  ^{2}$ the representation for the amplitude of the rare
$\pi^{0}\rightarrow e^{+}e^{-}$ process in the limit $q^{2}\rightarrow0$. It
is given in terms of the inverse moment of the transition pion form factor in
symmetric kinematics for space-like photon momenta. By using data of the CELLO
and CLEO Collaborations on the pion-photon transition form factor in the
obtained representation, we found the new lower bound for the decay branching
ratio which essentially improves the well-known unitary bound. Further
constraints follow from the results of OPE QCD\ correlating the pion
transition form factor in different kinematics as the change of characteristic
scales. These considerations allow us to reconstruct the full decay amplitude
and make predictions for the decay branching. A similar procedure is also
applied to the decays $\eta\rightarrow l^{+}l^{-}$. We compared our
predictions with the results obtained in various phenomenological approaches
and found that all of them are in agreement with our results. However, the
obtained prediction for the branching ratio $\pi^{0}\rightarrow e^{+}e^{-}$ is
3$\sigma$ below the recent KTeV measurement.

We are grateful to A.B. Arbuzov, M.V. Chizhov, J. Gasser, S.B. Gerasimov, N.
I. Kochelev, E.A. Kuraev, S.V. Mikhailov, O.V. Teryaev, Z.K. Silagadze for
helpful discussions on the subject of this work. We also acknowledge partial
support from the Heisenberg--Landau program and (AED) the Scientific School
grant 4476.2006.2.
%\bibliographystyle{plain}
%\bibliographystyle{prsty}
%\bibliographystyle{apsrev}
%\bibliographystyle{h-physrev3}
%\bibliography{Piee}

\end{document}